\documentclass[12pt]{iopart}
\usepackage{iopams}  
\usepackage{graphicx}

\newcommand{\myvec}[1]{\vec{#1}}
\newcommand{\vk}{{\myvec{k}}}
\newcommand{\vr}{{\myvec{r}}}
\newcommand{\vA}{{\myvec{A}}}
\newcommand{\vnabla}{{\myvec{\nabla}}}
\newcommand{\Ef}{\mathcal{E}}

\newcommand{\vEf}{\vec{\Ef}}

\newcommand{\ddt}{\frac{d}{dt}}

\renewcommand{\r}{\rangle}
\renewcommand{\l}{\langle}
\newcommand{\up}[1]{ ^{(#1)}}

\newcommand{\beq}{\begin{equation}}
\newcommand{\eeq}{\end{equation}}
\newcommand{\bea}{\begin{eqnarray}}
\newcommand{\eea}{\end{eqnarray}}

\begin{document}

\title[t-SURFF: two-electron photo emission]
{t-SURFF: Fully Differential Two-Electron Photo-Emission Spectra}

\author{Armin Scrinzi}

\address{Ludwig Maximilians Universit\"at, Theresienstrasse 37, 80333 Munich, Germany}
\ead{armin.scrinzi@lmu.de}
\begin{abstract}
The time dependent surface flux (t-SURFF) method is extended to single and double ionization
of two electron systems. Fully differential double emission 
spectra by strong pulses at extreme UV and infrared wave length are calculated using 
simulation volumes that only accommodate the effective range of the atomic binding potential
and the quiver radius of free electrons in the external field.
For a model system we find pronounced dependence of shake-up and non-sequential 
double ionization on phase and duration of the laser pulse. Extension to fully 
three-dimensional calculations is discussed. 
\end{abstract}

\maketitle

\section{Introduction}

Differential double photo-electron spectra and the corresponding ionic recoil momentum spectra 
testify of  dynamical correlation between the electrons.  
By sweeping extreme ultra violet (XUV) photon energies from below to above
the threshold for single-photon double ionization of the He atom one probes correlation in 
initial and final states. In wave length in the infrared (IR) range, momentum
distributions of recoil ions provide evidence for the importance of 
re-collision processes \cite{corkum93:simple-man}, where first one electron 
is ionized, which subsequently  is re-directed by the oscillating laser field 
into a collision with its parent
ion causing excitation and possibly detachment of the second electron.
The early observation of unexpectedly enhanced double ionization of 
helium in IR fields \cite{chaloupka03:double-ionization} is now generally 
ascribed to this mechanism. Experimental data on strong field IR photo-ionization 
is also available for many other atomic and molecular systems
and it was even proposed to use re-collision electron spectra for the analysis
of structure and dynamics of molecules \cite{meckel08:diffraction}.

For the XUV wave-length, theoretical and experimental questions have
matured, even if still under debate (see, e.g., \cite{Pazourek2011} 
for a recent contribution to the debate with ample references to 
theory and experiment).
At longer wave length, the large body of experimental data 
(see, e.g. 
\cite{lhuillier83_multiple_ionization,augst95:argon-3plus,mooshammer00:multiple_neon,
rudenko04:multi-electron,liu10:argon-neon}
and references therein) 
and the somewhat smaller range of theoretical models largely based on classical 
or semi-classical methods
(see, e.g., 
\cite{becker05:multiple-ionization,ye08:classical-finger,shvetsov-shilovski11:sequential,emmanouilidou11:double-prevalence} )
are all plagued by the almost complete lack of reliable theoretical 
verification, with the notable exception of a few very large scale simulations
of two-electron systems in strong fields \cite{awashti05:tdci,parker06:2e-cutoff,vanderhart06_floquet},
where, however, only in Ref.~\cite{parker06:2e-cutoff} the full two-electron dynamics is
treated for laser wavelength of 800 nm. There are several reasons for this 
striking absence of complete {\it ab initio} simulations of ionization of two-electron 
systems.
Firstly, even single ionization is non-trivial to compute, if the external fields
are non-perturbative. Roughly speaking, the effort for computing
single ionization grows with the 4th power of the wave length $\lambda^4$
due to the growth of peak momenta $\propto\lambda$, quiver amplitude of the
electron motion in the field $ \propto \lambda^2$ and the growth of minimum
pulse duration $\sim\lambda$ (see also the discussion in Ref.~\cite{tao12:ecs-spectra}).
When the effect of the field is perturbative, the situation for single ionization 
relaxes somewhat, as one basically only needs to know the initial neutral state and the 
single-electron stationary scattering solutions in the energy range of interest.
Although obtaining scattering solutions may be difficult, there is a well defined
procedure and the whole technology of electron-ion scattering theory available
to approach the problem.  
For double ionization, also in the perturbative regime the situation is more 
complex. The convenient partition into bound and singly ionized spectral eigenfunctions
cannot be continued to above the double-ionization threshold: 
eigenfunctions above the double-ionization threshold will in general
have, both, single- and double-ionization asymptotics.
For distinguishing single from double ionization we therefore invariably 
need the solutions at large distances. Even without the need for asymptotic
analysis, scattering with open double ionization channels is a challenging task.

Numerical simulations, in principle, can provide the full answer.
The asymptotic analysis is usually done by propagating the wave function 
until after the end of the pulse and then analyzing its remote
parts either in terms of momentum eigenfunctions, i.e. 
plane waves, or, when the tails of the ionic Coulomb potentials 
are not considered negligible, in terms of two-body Coulomb scattering
solutions. Three-body scattering solutions, which would obviate
a purely asymptotic analysis, unfortunately, are not accessible.
The by far largest part of the computational effort in these simulations goes 
into following the solution to large distances until the pulse is over and
analysis can begin. It may be irritating to think that a large effort is made
to simulate dynamics that is known exactly in the case of finite range potentials
or approximately at sufficient distances from all Coulomb centers: the free motion
of one or two electrons in a laser field.

In a preceding publication \cite{tao12:ecs-spectra} we have shown how, by absorbing
the wavefunction and recording of flux before absorption,
single particle photo-electron spectra can be computed using simulation
volumes that only contain the relevant range of the atomic potential 
and the electronic quiver motion in the field. 
In this so-called time-dependent surface flux (t-SURFF) method, asymptotic
information is accumulated during time propagation rather then drawn 
from the full wave function after the end of the pulse. The equivalence of
both approaches was proven in \cite{tao12:ecs-spectra} mathematically and by comparing
numerical results with literature. Depending on the
system parameters and accuracy requirements, absorption
radii for typical strong field setups can be kept as small as 20 Bohr radii
and as few as 90 linear discretization coefficients for the radial motion
can give better that 1\% accurate spectra over the complete spectral range.
For Coulomb systems, due to the long interaction,
larger radii of $\sim 100$ Bohr and 200 to 300 discretization points
may be needed.

In the present paper, we extend this approach to two-electron systems and
multi-channel single ionization as well as double ionization spectra. 
A numerical demonstration is provided using a two times one-dimensional 
model system. As first physical results, we find phase and pulse-duration
dependence of shake-up and double-ionization spectra, phenomena
that are consistent with the spatial asymmetry of very short 
laser pulses and the re-collision mechanism for shake-up and non-sequential double
ionization.  

\section{The t-SURFF method for two-electron systems}
We first briefly summarize t-SURFF for 
the single particle case. The basic requirement is 
that there is some radius $R_c$ beyond which the Hamiltonian 
reduces to an asymptotic one with known solutions. Let $H(t)$ denote
the time dependent Hamiltonian of our system and assume that there exists
an exactly solvable $H_v(t)$ that at large distances agrees with $H(t)$:
\beq
H_v(t)=H(t)\quad {\rm \,for\, } |\vr|>R_c\,{\rm \, and }\quad\forall t. 
\eeq
For a single particle in a laser field described in 
velocity gauge and a short range potential $V(\vr)\equiv0$ for $|\vr|>R_c$, 
$H_v(t)$ is the Hamiltonian for the free motion in the laser field 
\beq
H_v(t) = \frac12 [-i\vnabla-\vA(t)]^2,
\eeq
where $\vA(t)=-\int_{-\infty}^t \vEf(t') dt'$ for an electric dipole field $\vEf(t)$. 
Here and throughout we use atomic units $\hbar=m_e=e^2=1$
unless indicated otherwise ($m_e$  and $e$ denote electron mass and unit charge, respectively, 
the Bohr radius results as the atomic unit of length).
The TDSE with $H_v(t)$ has the Volkov solutions
\beq
\chi_\vk(\vr,t)=(2\pi)^{-3/2}e^{-i\Phi(\vk,t)}e^{i\vk\cdot\vr},
\eeq
i.e. plane waves times with time-dependence by the well-known Volkov phase
\beq
\Phi(\vk,t)=\frac12\int_{-\infty}^t dt' [\vk-\vA(t')]^2.
\eeq
Scattering describes the behavior of the time-dependent 
wave function $\Psi(\vr,T)$ at long times $T$ and large distances $|\vr| > R_c$.
We choose $T$ and $R_c$ large enough such that at $T$ the pulse is over
\beq\label{eq:hc_T}
H_v(t)=-\frac12 \Delta \quad{\rm for}\ t>T
\eeq
and the wave function has split into its bound and scattering parts 
\beq
\Psi(\vr,T) = \Psi_{\rm b}(\vr,T) + \Psi_{\rm s}(\vr,T)
\eeq
with the properties
\bea\label{eq:psi-b}
\Psi_{\rm b}(\vr,T)\approx 0 &\quad{\rm for}\ |\vr|\ge R_c \\
\label{eq:psi-s}
\Psi_{\rm s}(\vr,T)\approx 0&\quad{\rm for}\ |\vr|\le R_c.
\eea
The approximate sign in (\ref{eq:psi-b}) refers to the tails of any bound state function,
which decay exponentially with increasing $R_c$.
The approximate sign in (\ref{eq:psi-s}) refers to the fact that electrons with very low energies $\vk^2/2\sim 0$ 
may not have passed $R_c$ at time $T$. 
We only need to analyze $\Psi_s(\vr,T)$ in terms of asymptotic functions $\chi_\vk(\vr,T)$. 
As $\Psi_s(\vr,T)$ vanishes inside the radius $R_c$, we can multiply the full $\Psi(\vr,T)$ by the 
function
\beq\label{eq:theta}
\theta(\vr,R_c) = \left\{\begin{array}{ll} 0&{\rm\,for\,} |\vr|<R_c\\  1&{\rm\,for\,} |\vr|\ge R_c\end{array}\right.
\eeq  
write the emission amplitude for photo-electron momentum $\vk$ as
\begin{equation}\label{eq:b-amplitude}
b(\vk,T) = \l \chi_\vk(T)|\theta(R_c)|\Psi(T)\r.
\end{equation}
For obtaining a time integral over a surface flux, we write (\ref{eq:b-amplitude}) 
as an integral of the time derivative and use the fact on the support of $\theta(R_c)$
both, $\chi_\vk(\vr,t)$ and $\Psi(\vr,t)$  evolve by the same Hamiltonian $H_v(t)$.
We obtain
\beq\label{eq:surfint1}
\l \chi_k(T) |\theta(R_c)| \Psi(T)\r
=i\int_0^T dt \l \chi_k(t) |[H_v(t),\theta(R_c)]| \Psi(t)\r
\eeq
The commutator vanishes everywhere except on $|\vr|=R_c$:
the asymptotic information is obtained by integrating the time dependent flux 
through a surface at finite distance $R_c$.
For the further discussion of the single-particle case and numerical examples 
with short-range and Coulomb potentials we refer to Ref.~\cite{tao12:ecs-spectra}. 

\subsection{Single-ionization into ionic ground and excited state channels}
\label{sec:single}

The simplest extension of the single electron method is for
computing single ionization into ground and excited state ionic channels. 
The complication is that in presence of a strong field the ionic state can differ from
the field free ionic state due to polarization and one must make sure to count 
flux passing the surface into the correct ionic channel. In this section we consider only 
the lowest ionic states that remain bound and do not contribute to double ionization.

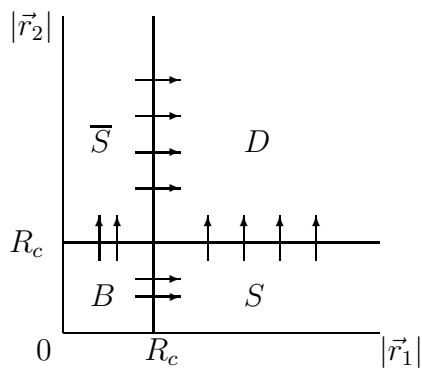
\begin{figure}
\begin{center}
\setlength{\unitlength}{1.2mm}
\begin{picture}(35,35)
  \put(-6,33){$|\vr_2|$}
  \put(-6,9){$R_c$}
  \put(0,0){\line(0,1){35}}
  \put(3,20){$\overline{S}$}
  \put(10,0){\line(0,1){35}}
  \put(3,3){$B$}
  \put(-3,-3){0}
  \put(20,20){$D$}
  \put(0,10){\line(1,0){35}}
  \put(20,3){$S$}
  \put(0,0){\line(1,0){35}}
  \put(9,-3){$R_c$}
  \put(35,-3){$|\vr_1|$}

  \put( 8,4){\vector(1,0){5}}
  \put( 8,6){\vector(1,0){5}}
  \put(16,8){\vector(0,1){5}}
  \put(20,8){\vector(0,1){5}}
  \put(24,8){\vector(0,1){5}}
  \put(28,8){\vector(0,1){5}}

  \put(4, 8){\vector(0,1){5}}
  \put(6, 8){\vector(0,1){5}}
  \put(8,16){\vector(1,0){5}}
  \put(8,20){\vector(1,0){5}}
  \put(8,24){\vector(1,0){5}}
  \put(8,28){\vector(1,0){5}}

\end{picture}\vspace*{2ex}
\caption{\label{fig:partition}
Partitioning of coordinate space into bound ($B$), singly ionized ($S,\overline{S}$) and doubly 
ionized $D$ regions. Single and double photo electron spectra are obtained by integrating the 
flux across the boundaries between the regions. 
}
\end{center}
\end{figure}

For decomposing coordinate space into bound and asymptotic 
regions we define on both coordinates $\vr_1$ and $\vr_2$ 
projector functions $\theta_1(\vr_1,R_c)$ and  $\theta_2(\vr_2,R_c)$, respectively, 
as in the single particle case (\ref{eq:theta}). Again picking sufficiently large times
$T$ and a sufficiently large surface radius $R_c$ we can partition the 
wave function $\Psi(T)$ into its bound, singly ionized, and doubly ionized parts 
(see Figure~\ref{fig:partition})
\bea
\Psi(T)&=&(1-\theta_1)(1-\theta_2)\Psi(T) 
\\&&+ \theta_1(1-\theta_2)\Psi(T) 
+ (1-\theta_1)\theta_2\Psi(T) +
\theta_1\theta_2\Psi(T)
\nonumber\\
&:=&\Psi_b(T)+\Psi_s(T)+\Psi_{\overline{s}}(T) + \Psi_d(T)  
\eea
Here and in the following we suppress the arguments $\vr_1, \vr_2$ and $R_c$ of $\theta_1$ and $\theta_2$.
Note that the assignment of singly and doubly ionized character to the different regions
is asymptotically exact: 
the error can be made arbitrarily small for any specific solution $\Psi(T)$
by choosing sufficiently large $T$ and $R_c$.

The single ionization Hamiltonian
\beq\label{eq:hsingle-ion}
H_s(t) = H_v(t)\otimes H_{ion}(t)
\eeq
agrees with the exact Hamiltonian on the support of $\theta_1(1-\theta_2)$ (region $S$ in Figure~\ref{fig:partition}).
The corresponding Hamiltonian on $\overline{S}$ is obtained by 
 particle exchange.
Channel solutions  $\chi_{c,\vk}(\vr_1,\vr_2,t)$ for the TDSE on $S$
\beq
i\ddt \chi_c(\vr_1,\vr_2,t)=H_s(t)  \chi_c(\vr_1,\vr_2,t)
\eeq
have the form
\beq
 \chi_{c,\vk}(\vr_1,\vr_2,t)= (2\pi)^{-3/2}e^{-i\Phi(t)}e^{i\vk\vr_1}\otimes \phi_c(\vr_2,t).
\eeq
Here $\phi_c(\vr_2,t)$ solves the ionic TDSE  
\beq
i\ddt \phi_c(\vr_2,t) =  H_{ion}(t) \phi_c(\vr_2,t).
\eeq
Rather than an initial we use a {\em final} 
condition that $\phi_c(\vr_2,t)$ for $t>T$ should be the desired final ion state
(we need to solve the TDSE backward in time).
The channel solution on the support $\overline{S}$ of the particle exchanged projector 
$(1-\theta_1)\theta_2$ is $\overline{\chi}_{c,\vk}(\vr_1,\vr_2,t)=\chi_{c,\vk}(\vr_2,\vr_1,t)$.

With the spectral amplitude
\beq
b(\vk,c,T)=\l \chi_{\vk,c}(T) | \theta_1(1-\theta_2) |\Psi_{\rm s}(T)\r.
\eeq 
the probability density for finding at time $T$ an electron with momentum $\vk$ in ionic channel $c$ 
is
\beq
\sigma(\vk,c,T)=2|b(\vk,c,T)|^2.
\eeq
The factor 2 arises from adding the two identical exchange symmetric contributions.
For converting this integral into a time integral over a surface we make the simplifying
assumption that the ionic solution never leaves the bound area 
\beq\label{eq:zeroflux}
\phi_c(\vr_2,t) \approx 0 {\quad\rm for\,} |\vr|>R_c \quad{\rm and\quad}\forall t.
\eeq
This is the precise version of the assumption that the ionic states considered does not get
further ionized. The approximate sign refers to the fact that again there is
always an exponential tail reaching to arbitrary distances and further that interaction with any pulse will
lead to a small amount of ionization. Using (\ref{eq:zeroflux}) we neglect the flux between
the singly ionized regions $S,\overline{S}$ and the doubly ionized region $D$ at all times.
Then, using the same procedure as for a single particle we obtain
\beq\label{eq:channel}
b(\vk,c,T)=i\int_{-\infty}^T dt \l \vk, t |[H_v(t),\theta_1]| \psi_c(t)\r
\eeq 
where $\psi_c(\vr_1,t)$ is the channel projected wave function
\beq
\psi_c(\vr_1,t):=\int d^3r_2 \phi^*_c(\vr_2,t)\Psi(\vr_1,\vr_2,t).
\eeq
For evaluating the integral (\ref{eq:channel}) we only need to know 
values and radial derivatives of $\psi_c$ on the surface $|\vr|=R_c$.  

For the computation it means solving the full two-electron problem up
to time $T$ and up to radius $R_c$. Beyond $R_c$ one can absorb all amplitudes. 
In addition, for each channel $c$, we need to solve one 
single electron problem up to the same time and radius (which is usually a 
much simpler task). 

\subsection{Double ionization spectra}

When double ionization occurs, flux passes from the bound region $B$ through 
the singly ionized regions $S$ or $\overline{S}$ into the doubly ionized 
region $D$. Our naming of the areas is suggestive but does not 
imply any bias as to an actual state that the electrons occupy
within any of these areas. It is unsubstantial for the present discussion 
whether some intermediate ionic bound state is occupied by one of the electrons 
in $S$ or $\overline{S}$ (sequential ionization) or whether both electrons
must be considered unbound. 
The sole purpose of the partitioning is to have well-defined surfaces 
outside the ranges of the respective potentials where we will integrate
fluxes.

For double ionization there is one obvious limitation of the discussion
so far: on the line $|\vr_1-\vr_2|=a$ the electron-electron interaction
is constant and not negligible for small $a$. 
This problem is not related to the long-range Coulomb potential,
rather it is a general problem for any multi-particle breakup, which is why
break-up processes are more complex than single particle 
scattering. Within the framework of the present approach the problem can be 
solved  for short-range electron-electron repulsion without making approximations,
which will be discussed below. 
A pragmatic solution has been effectively employed in many earlier publications,
which is to neglect electron-electron repulsion at large distances from 
the nucleus. This is what using any projection onto products of single electron
states implies, be it Coulomb scattering waves or plane waves (both approaches are
discussed, for example, in \cite{feist08:helium}).
Sensitivity to this approximation can be tested by varying the distance
from the nucleus where the projection starts.

As the pragmatic solution was found to work well in many cases, 
we make this approximation explicit in the present paper 
by smoothly turning off all potentials including the electron-electron interaction
before the surface radius $R_c$. In that case we can always use the free
(Volkov) Hamiltonian $H_v(t)$ beyond $R_c$.
 
By our assumptions, in the region $|\vr_1|\geq R_c$, the Hamiltonian
is identical to $H_s(t)$, Eq.~(\ref{eq:hsingle-ion}), which motivates the ansatz
\beq\label{eq:psi_theta1}
\theta_1\Psi(\vr_1,\vr_2,t)=\int d^3k \sum_n \chi_\vk(\vr_1,t)\xi_n(\vr_2)b'(\vk,n),
\eeq
with the Volkov solutions $\chi_\vk(\vr_1,t)$ on coordinate $\vr_1$ 
and an expansion into a
time-independent, complete, but otherwise arbitrary set of functions 
$\xi_n(\vr_2)$ on $\vr_2$.
Using orthogonal projection onto the expansion functions $\chi_{\vk_1}$ and $\xi_n$, 
the coefficients $b'(\vk_1,n,t)$ are obtained as
\beq\label{eq:inverse}
b'(\vk_1,n,t)=\int d^3k_1' q_\theta(\vk_1,k_1',t) b(\vk_1',n,t) 
\eeq
with 
\beq
b(\vk,n,t)=\int_{-\infty}^\infty d^3r_1 \chi^*_{\vk_1'}(\vr_1,t) \theta_1\int_{-\infty}^\infty \xi^*_n(\vr_2) \Psi(\vr_1,\vr_2,t).
\eeq
The integral (\ref{eq:inverse}) over $q_\theta$ accounts for the fact that the plane waves 
are not $\delta$-orthonormal when the integration is restricted by $\theta_1$:
the inverse overlap is defined by
\beq\label{eq:kthetak}
\int d^3k' q_\theta(\vk,\vk',t) \l \vk', t|\theta_1 |\vk'' ,t\r =\delta^{(3)}(\vk-\vk''). 
\eeq
For notational brevity, we assume here that the basis functions $\xi_n$ are
orthonormal. The time-derivative of the $b(\vk_1,n,t)$ is
\bea
i\ddt b(\vk_1,n,t)&=& 
\l \chi_{\vk_1}(t) | \theta_1 \l \xi_n | H_2(t) | \Psi(t)\r\r   
\nonumber\\ && -\l \chi_{\vk_1}(t)|[H_v(t),\theta_1]\l \xi_n |\Psi(t)\r\r
\eea
By the double right bracket $\r\r$ we emphasize that integration is over both coordinates $\vr_1$ and $\vr_2$.
Inserting the representation (\ref{eq:psi_theta1}) into the first term
we obtain an inhomogeneous equation for the $b(\vk_1,n,t)$: 
\bea
i\ddt b(\vk_1,n,t)  
&=& \sum_m \l \xi_n | H_{ion}(t)|m\r b(\vk_1,m,t) 
\nonumber\\ &&
-\l \vk_1,t|[H_v(t),\theta_1]\l \xi_n |\Psi(t)\r\r
\label{eq:b-evolution}\eea
where we have used (\ref{eq:kthetak}). The inhomogeneity is the
flux through the surface $|\vr_1|=R_c$. Initial conditions are $b(\vk_1,n)\equiv0$,
i.e.  no electrons outside $R_c$.
We write the double ionization amplitude 
\beq
b(\vk_1,\vk_2,T)=\l \chi_{\vk_2}(T) | \l \chi_{\vk_1}(T) | \theta_2 \theta_1 | \Psi(T)\r\r
\eeq
and use one more time the conversion to integrals over surface flux
\bea
\lefteqn{b(\vk_1,\vk_2,T)=\l \chi_{\vk_2}(T) | \l \chi_{\vk_1}(T) | \theta_2 \theta_1 | \Psi(T)\r\r}\\
&=&\int^T dt \ddt \l \chi_{\vk_2}(t) | \l \chi_{\vk_1}(t) | \theta_2 \theta_1 | \Psi(t)\r\r \\
&=:&i\int^T dt [B(\vk_1,\vk_2,t)+ \overline{B}(\vk_1,\vk_2,t)].
\eea
The two terms $B$ and $\overline{B}$ are related by exchange symmetry
\beq
B(\vk_1,\vk_2,t)=\overline{B}(\vk_2,\vk_1,t)=\l \chi_{\vk_2}(t)|  \l \chi_{\vk_1}(t) [H_v(t),\theta_2]\theta_1 | \Psi(t)\r\r
\eeq
For computing $B$, we only need to know $\theta_1\Psi(t)$, for which we insert the
representation (\ref{eq:psi_theta1}) to obtain
\bea
B(\vk_1,\vk_2,t)&=& \sum_m \l \chi_{\vk_2}(t)|[H_v(t),\theta_2]|\xi_m\r b(\vk_2,m,t)
\eea
The inverse overlap $q_\theta(\vk_1,\vk_1')$, (\ref{eq:kthetak}), cancels with the overlap 
integrals and never needs to be evaluated explicitly.

$B(\vk_1,\vk_2,t)$ is the contribution to the double ionization spectrum passing at time $t$ through 
surface $|\vr_2|=R_c$ from region $S$ into $D$. In $S$, the first electron is already detached
and has a fixed canonical momentum $\vk_1$ that is carried into the double-ionized region $D$.
$\overline{B}(\vk_1,\vk_2,t)$ is the alternate contribution going through region $\overline{S}$.

\subsection{Computational remarks}

The substantial gain of the method is that, rather than computing the full solution
in region $D$ and then analyzing it, we only need to integrate the flux through the surface separating 
$S$ from $D$. In the direction parallel to that surface the wave function is represented
in terms of the free solutions $\chi_{\vk_1}(\vr_1,t)$, where we do not need to 
expand the wave function completely, but can restrict propagation to the momenta 
$\vk_1$ that we are interested in. For each $\vk_1$ we need to solve equation (\ref{eq:b-evolution}),
which is an ionic TDSE with an additional source term (the flux entering $S$ from the 
bound region $B$). Although the derivation may appear complex, the implementation
of the procedure can be done by the following simple algorithm
\begin{itemize}
\item Set up $b(\vk_1,n,t)$ on a $\vk_1$ grid of the desired density and initialize to 0.
\item Get time grid points $t_i$ with a time step that resolves the 
fastest amplitude oscillations for the desired energy range $\Delta t< 2\pi/E_{max}$ 
\item In a loop through all times $t_i$, get the surface terms from file or 
from a simultaneous calculation of $|\Psi(t_i)\r\r$ and use (\ref{eq:b-evolution}) to advance to $b(\vk_1,n,t_i)$. 
\item Add the contribution from region $S$  to the spectral amplitude
\beq
s(\vk_1,\vk_2,t_i)=s(\vk_1,\vk_2,t_{i-1})+\Delta t B(\vk_1,\vk_2,t).
\eeq
\end{itemize}
The contribution $\overline{s}$ from region $\overline{S}$ is obtained from $s$ by particle
exchange. The total spectral amplitude is the sum of both contributions 
\beq
b(\vk_1,\vk_2,T)=s(\vk_1,\vk_2,T)+\overline{s}(\vk_1,\vk_2,T)=s(\vk_1,\vk_2,T)+s(\vk_2,\vk_1,T)
\eeq
The asymptotic value --- the spectral amplitude --- is attained at times $T$ when the 
flux through the surfaces becomes negligible.

The approach can only be successful, when absorption does not significantly distort 
the solution at the integration surfaces. The ``infinite range exterior complex scaling'' (irECS)
absorber was shown in Refs.~\cite{tao12:ecs-spectra,scrinzi10:irecs} to provide traceless absorption
over a very wide energy range at low computational cost. It outperforms standard complex 
absorbing potentials by several orders of magnitude in accuracy. If needed, irECS 
can be pushed to full machine precision using not more than 20 discretization coefficients
per coordinate in the absorbing region $|\vr|>A_0$.
Absorption can begin at any $A_0\geq R_c$. More mathematical and numerical detail 
and ample numerical examples using irECS can be found in Refs.~\cite{tao12:ecs-spectra,scrinzi10:irecs}.

In general discretization errors for two electrons are similar to those for 
a single-electron system with the same ionization potential. This has the practical
advantage that a suitable discretization can be determined from the single-electron problem. 
Only a few consistency checks on specific two-electron observables need to be performed 
for the computationally heavier two-electron calculation.

\subsubsection{General single ionization spectra}
Once we have obtained the $b(\vk_1,n,T)$ we can reconstruct the wave function for $|\vr_1|>R_c$.
In particular, the amplitude for single ionization spectra for any ionic state $\phi_c$ is
\beq\label{eq:channel2}
b(\vk,c)=\sum_n b(\vk,n) a\up{c}_n,
\eeq
where $a\up{c}_n$ are the expansion coefficients of $\phi_c$ with respect to the basis
functions $\xi_n$:
\beq
\phi_c(\vr)=\sum_n \xi_n(\vr) a\up{c}_n.
\eeq
This approach is not limited to non-ionizing $\phi_c$, as it analyzes
ionic population at time $T$ after the end of the pulse. For non-ionizing $\phi_c$, it 
is an alternative to the single-ionization procedure above. Note that 
here we need to solve the inhomogeneous ionic problem (\ref{eq:b-evolution}) 
for each photo-electron momentum $\vk$. The total spectrum is a linear
combination of the individual contributions from each $n$. Where it is applicable, 
the advantage 
of the single-ionization procedure of section \ref{sec:single} twofold: firstly, 
for each final $\phi_c$ one needs to solve only one ionic TDSE  and 
compute values and derivatives of the channel surface function $\psi_c$. 
The complete spectrum can be obtained by 
time-integration with little numerical effort. Secondly, as one directly obtains
the channel spectrum without intermediated decomposition and final resummation of
the wave function, results are in general more robust numerically.

\section{Numerical demonstration of the method}

\subsection{The one-dimensional two-electron model system}

As even with the simplifications by t-SURFF the solution of the
full three-dimensional two-electron problem remains a very large scale
computational task, we use for demonstration purposes the standard 
one-dimensional two-electron model Hamiltonian
\beq
H(t) = \sum_{\alpha=1,2} 
\frac12 \left[-i\frac{\partial}{\partial x_\alpha} - A(t)\right]^2
-\frac{2M(x_\alpha)}{\sqrt{x_\alpha^2+1/2}} + \frac{M(x_1)M(x_2)}{\sqrt{(x_1-x_2)^2+0.3}}.
\eeq
For making all potentials strictly finite range, we have
chosen a ``truncation radius'' $C_p$ and a truncation function $M(x)$ that 
is $\equiv1$ up to $|x_\alpha|=C_p-5$ and goes differentiably smoothly to 0 at 
$|x_\alpha|=C_p$. Where not indicated otherwise we use $C_p=20$.
The screening factor of $1/2$ in the ionic potential was chosen for esthetic reasons,
as then the exact ionic ground state (without potential truncation) is $E_0=-2$.
The first excited state occurs at $E_1=-0.93$, substantially stronger bound than the first 
excited $He^+$ energy of $-1/2$. 
With the electron-electron screening of $0.3$ one obtains an ionization potential 
of $0.88$, which we consider a fair approximation of the actual He ionization potential of
$0.90$.

For all computations we use a finite elements discretization where orders between 
8 and 20 where used for convergence studies. For the present purposes we found 
order 8 or 10 sufficient. A deeper discussion of the finite element method for 
irECS calculations can be found in \cite{scrinzi10:irecs}. The irECS radius $A_0$
which marks the beginning of absorption was varied between 20 and 100 atomic units.
Results do not depend on $A_0$  on the level of accuracy shown. 

In all calculations below we use $\cos^2$ pulse shapes defined in terms of the 
vector potential 
\beq
A(t) = \frac{\Ef_0}{\omega}\cos^2(\frac{\pi t}{2T_{FWHM}})\sin(\omega t + \phi_{CEO}),
\eeq
where $T_{FWHM}$ is the full-width-half-maximum of the envelope of the vector potential, 
$\omega$ is the laser central photon energy, and $\phi_{CEO}$ the carrier-envelope offset phase.
The peak field amplitude $\Ef_0$ is related to the pulse 
peak intensity $I$ by $\Ef_0= \sqrt{2I}$. 
By construction, the electric field $\Ef(t)=-\ddt A(t)$ has no zero frequency 
component, which is important when studying effects of $\phi_{CEO}$ with very short
pulses. 
For a ``cosine pulse'' $\phi_{CEO}=0$ the peak of the electric field approximately 
coincides with the maximum of the pulse envelope with minor deviations due to the 
derivative of the envelope.

\subsection{Two photon double ionization in the extreme ultraviolet}
\label{sec:xuv}

In Ref.~\cite{Pazourek2011} perturbative two-photon double ionization
of He by short XUV pulses with photon energies between the two- and single-photon
ionization thresholds
was investigated and a remarkable universal description of the process
was found. The findings were confirmed by numerical solutions of the 
fully three-dimensional two-electron TDSE. Pulse durations of 
$T=4.5\,fs$ and photon energies $\omega$ between 42 and 80 eV were used.
The perturbative regime at $\omega=42\,eV$ extends up to intensities
near $I=10^{16}W/cm^2$, where one first observes sizable deviations from the
perturbative scaling $\propto I$ at the single-photon single-ionization peak
and $\propto I^2$ for shake-up and double ionization.

\begin{figure}
\includegraphics[height=14cm,angle=-90]{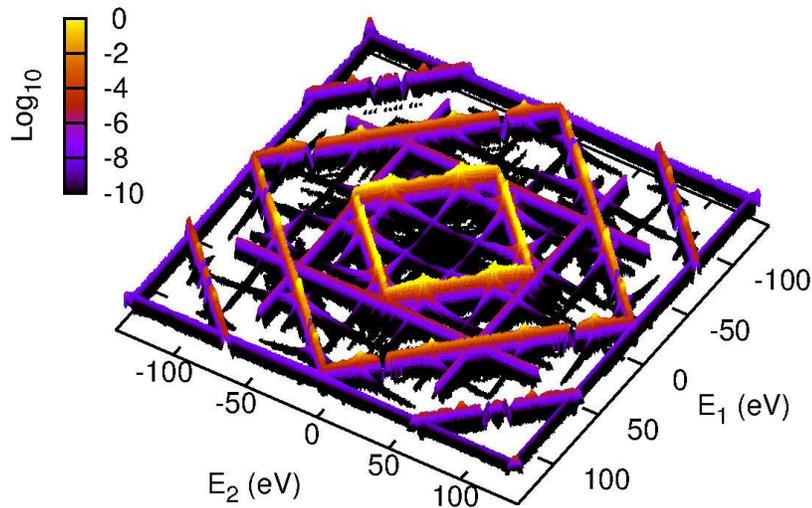}
\caption{\label{fig:xuv-overview}
Double electron emission by a 4.5 fs XUV pulse at photon energy $\omega=70 eV$.
with peak intensity $10^{16}W/cm^2$. 
Negative energies indicate emission to $(-\infty,-R_c]$. Straight lines at
45 degrees with constant $E_1\pm E_2$ indicate correlated processes.
The first 3 double-electron ridges are visible. 
Lines parallel to the energy axes are artefacts due to exponential tails of 
excited ionic bound states that reach beyond $R_c=40$. Their peak values 
lie below $\lesssim 10^{-6}$ of the peak signal.
Black color ($\leq 10^{-10}$)
approximately indicates numerical noise.
}
\end{figure}

As an example, Figure \ref{fig:xuv-overview} shows the two-electron spectrum
for photon energy $\omega=70\,eV$ at peak intensity $10^{16}W/cm^2$. Three two-electron
at energies $|E_1|+|E_2|=n\omega+E_0$ are clearly visible. With ground state
energy $E_0=-2.88\,au\approx-78\,eV$ the first peak appears for n=2-photon 
ionization. The ridges parallel to the energy axes are artefacts of the method:
there is significant shake-up and the exponential tail of the excited ionic 
states reaches into region $D$. Note, however, that with the choice of $R_c=40$ for the 
present calculation, the ridges are suppressed by at least a factor $10^{-6}$ relative
to the surrounding signal. Even stronger suppression can be obtained by further 
increasing $R_c$ or, alternatively and with slower convergence rate, propagating 
to longer times. If the ionic states are known accurately, their effect can be
completely removed by projection (for a detailed discussion of the procedure 
see Ref.~\cite{tao12:ecs-spectra}).

\begin{figure}
\includegraphics[height=10cm,angle=-90]{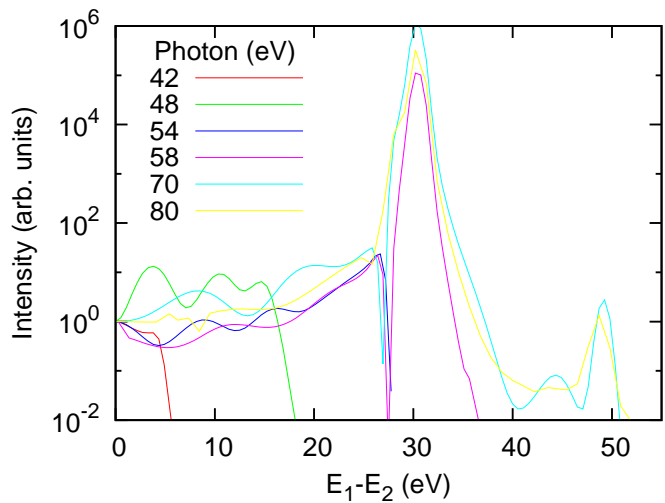}
\caption{\label{fig:xuv-lineout}
Two-photon double ionization for photon energies 42, 54, 58, 70, and 80 eV.
Lineouts are taken at the respective lowest energy ridges with
total energies $|E_1|+|E_2|=6,18,30,38,62$ and $82\,eV$, respectively.
Curves are normalized to 1 at $E_1-E_2=0$.
}
\end{figure}

In Figure~\ref{fig:xuv-lineout} we show line-outs of the two-photon two-electron energy
ridges for photon energies between 42 and 80 eV. The curves are converged
within the resolution of the plots with respect to 
$T$, $R_c$ and spatial discretization. Although the general characteristics
are determined by the single-excitation resonances as in  \cite{Pazourek2011}, for the 
present system we do not reproduce the universal behavior found there for the 
angle-integrated spectra. The likely reason are shake-up interferences that
also modify the the angular distributions in the 3D case \cite{Pazourek2011}.

Even at the present benign, nearly perturbative parameters, 
photo-electrons with energies of about 
$E_{\varphi}\sim 3 au\approx 80\, eV$ travel to a distance 
$r_{max}\sim 2T_{FWHM} \sqrt{2 E_{\varphi}/m_e}\approx 900\,au$ during the duration of the pulse. 
To correctly represent the corresponding momenta one needs a grid spacing of 
$\Delta r \lesssim 2\pi/ \sqrt{2E_{max}}\approx 2.5$,
which results in at least about 400 radial discretization points in a numerical calculation.
In practice, finer grid spacings need to be used with a correspondingly larger
number of discretization points.
In addition, for asymptotic analysis of the wave function, one may need to propagate
further for a certain period of time after the pulse, leading to further
increase in the required box radius $r_{max}$ proportional to propagation time.

For comparison, with t-SURFF we obtain converged results using only $49$ discretization
coefficients on the positive half axis $[0,\infty)$. The total number of discretization points
per coordinate is twice as large because of the negative half-axis, which would correspond to a different
angular direction in a three-dimensional calculation. 32 out of the 49 points were used on the 
interval $[0,20]$, i.e. an effective grid spacing of $\sim 20/32=0.625$, which sets a theoretical 
limit for the maximum photo-electron energy of $\lesssim 50\,au \approx 1300\,eV$. The remaining
17 points were used for irECS absorption. The t-SURFF simulation volume radius $R_c$ 
is independent of pulse duration.

\subsection{Single ionization and shake-up by an IR pulse}

Possibly the most elementary correlated process is shake-up: after forcefully
removing a single electron, the remaining electrons rearrange not exclusively
into the ionic ground state, but a fraction goes into excited states. The exact
distribution of excited states strongly depends on system parameters. 
Shake-up by very strong, short infrared pulses has been observed recently using the 
``attosecond transient absorption spectroscopy'' \cite{goulielmakis10:shake-up}, which can monitor
the time-evolution of an excited ionic state after IR ionization.

Within the present simplified model we can only demonstrate basic
qualitative features of shake-up in IR ionization. We have studied
dependence on the carrier-envelope offset phase $\phi_{CEO}$ and on pulse duration.
We use an intensity of $2\times10^{14}W/cm^2$ and $\omega=0.057 au$ (corresponding
to wave length 800 nm). Converged results were obtained using 
the same discretization parameters as in the XUV case discussed above.

\begin{figure}
\includegraphics[height=12cm,angle=-90]{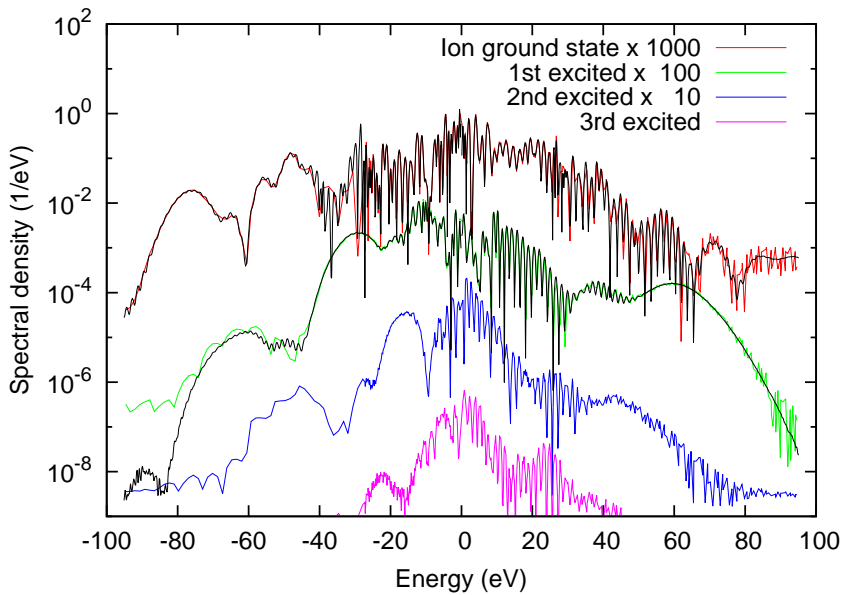}
\caption{\label{fig:ionic_channels}
Single photo-electron emission spectra in the lowest 4 ionic channels.
Negative energies indicate emission to $(-\infty,-R_c]$. 
Spectra are calculated using Eq.~(\ref{eq:channel2}).
In addition, for ground and first excited ion channel, curves computed by Eq.~(\ref{eq:channel}) are
shown (black lines). The curves are scaled for better visibility.
}
\end{figure}

In Figure~\ref{fig:ionic_channels} photo electron spectra for the ground and first excited
states calculated by formula (\ref{eq:channel}, smooth line) and, alternatively, by Eq.~(\ref{eq:channel2}, 
coarser energy grid) for a single cycle cosine pulse: $T=2\pi/\omega$ and
$\phi_{CEO}=0$. The spectra show some generic short pulse features: pronounced
asymmetry and absence of individual photo-electron peaks at energies, where emission
occurs only during a single laser half-cycle. Emission to the
left is lower, as the field amplitude to the left is smaller.

Clearly, the spectral structure is highly sensitive to $\phi_{CEO}$. However,
also the total shake-up yield is strongly $\phi_{CEO}$ dependent. 
Table~\ref{tab:shake-up} list the yields in the ground and first excited ionic
states for different phases and pulse durations. While shake-up is strongly suppressed 
for a single cycle cosine pulse, it increases to a sizeable $\sim10\%$ of the ground
state ionic channel for $\phi_{CEO}=3\pi/4$.
We tentatively ascribe this fact to a recollision mechanism for shake-up.
The fact that longer pulses with 2 and 3 optical cycles in general have
shake-up fractions comparable to the  $\phi_{CEO}=3\pi/4$ is consistent with 
this hypothesis. We abstain from a more detailed analysis of the phenomenon 
for the present model, as due to the absence of transverse wave-packet spreading,
recollision effects are greatly exaggerated in one dimensional models.

\begin{table}
\begin{tabular}{cc|ccc}
$T_{FWHM}$&$\phi_{CEO}$ & $Y_0$ & $Y_1$ & $Y_1/Y_0$\\
\hline
1 opt.cyc.&0&$2.26\times10^{-7}$&$6.27\times10^{-10}$&$2.77\times10^{-3}$\\
&$\pi/4$&$2.18\times10^{-7}$&$1.35\times10^{-8}$&$6.19\times10^{-2}$\\
&$\pi/2$&$8.88\times10^{-8}$&$1.35\times10^{-8}$&$1.15\times10^{-1}$\\
&$3\pi/4$&$1.38\times10^{-7}$&$2.60\times10^{-9}$&$1.88\times10^{-2}$\\
2 opt.cyc.&0&$2.36\times10^{-7}$&$1.16\times10^{-8}$&$4.91\times10^{-2}$\\
3 opt.cyc.&0&$1.44\times10^{-7}$&$3.18\times10^{-9}$& $2.21\times10^{-2}$\\
\end{tabular}
\caption{\label{tab:shake-up} Shake-up as a function of $\phi_{CEO}$. Total yield in 
the ionic ground $Y_0$ and first excited $Y_1$ states and ratio $Y_0/Y_1$.}
\end{table}

\subsection{Double ionization spectra generated by  an IR pulse}

\begin{figure}
\includegraphics[height=10cm,angle=0]{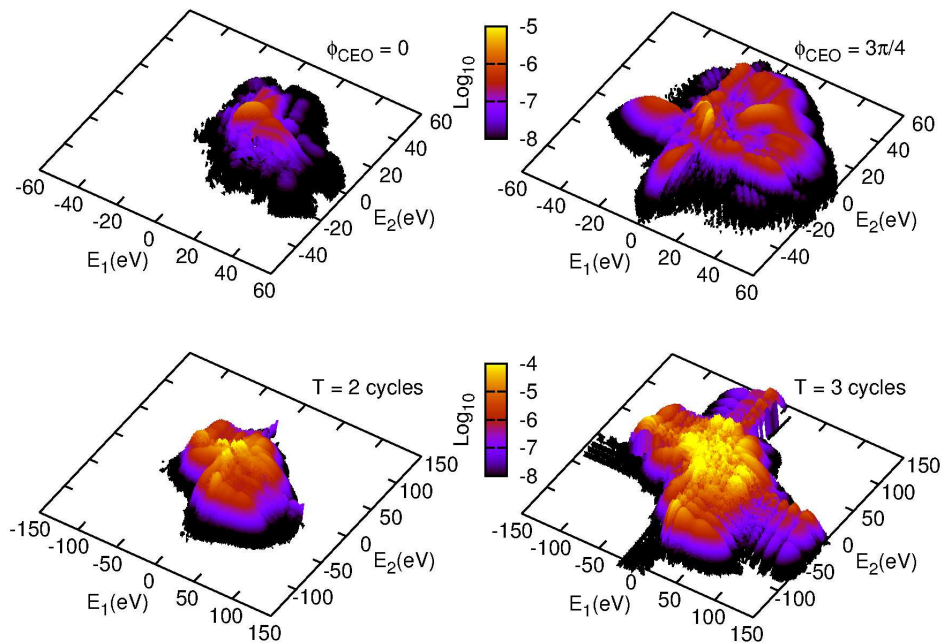}
\caption{\label{fig:overview}
Overview of double ionization spectra for laser wave-length 800 nm and peak intensity
$2\times10^{14}W/cm^2$. Upper row pulse duration $T=1$ optical cycle with $\phi_{CEO}=0$ (left)
and $3\pi/4$ (right). Lower row $\phi_{CEO}=0$ with $T=2$ (left) and 3 (right).
Negative energies indicate emission to $(-\infty,-R_c]$.
}
\end{figure}

Figure \ref{fig:overview} gives an overview of doubly-differential photo emission spectra
obtained for different $\phi_{CEO}$ and pulse durations up to three optical cycles
$T_{FWHM}=6\pi/\omega\,au\approx 7.5\,fs$. 
The calculations were performed using a slightly more accurate discretization
with 60 points on the half-axis $[0,\infty)$.

As to be expected, we see pronounced asymmetries and 
strong effects of $\phi_{CEO}$ for the shortest pulses. While the single-cycle 
cosine pulse essentially only shows weak and unidirectional emission of both electrons,
backward emission and significantly higher yields arise at $\phi_{CEO}=3\pi/4$.
Most likely, the responsible mechanism is re-collision.
However, due to the limitations of the one-dimensional model,
we defer a more profound discussion 
of the effects to a future fully three-dimensional calculation.

\subsubsection{Dependence on the potential cutoff}

\begin{figure}
\includegraphics[height=12cm,angle=-90]{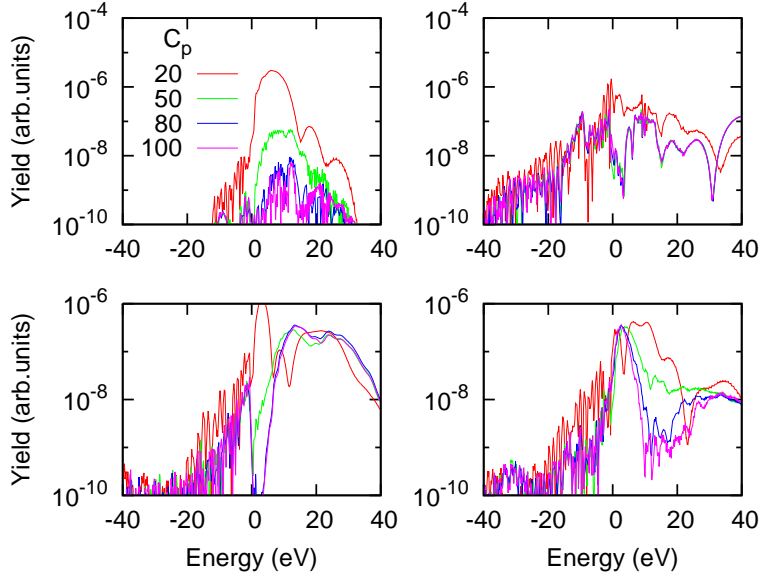}
\caption{\label{fig:rp-dependence}
Dependence of spectra on the potential cutoff $C_p=20,50,80$ and $100$
for a 800 nm single-cycle cosine pulse with peak intensity $2\times10^{14}W/cm^2$.
Top left: single ionization in the first shake-up channel $\sigma_1(E)$,
top right: double ionization with equal energies $\sigma(E,E)$,
bottom: lineouts of the double ionization signal $\sigma(E,E_1)$ for
energies $E_1=3\,eV$ (left) and $13\,eV$ (right).
}
\end{figure}

By truncating all potentials at $C_p=20\, au$, the Volkov Hamiltonian $H_v(t)$
is exact outside $R_c$ and all errors of t-SURFF are either due to discretization
or absorption. As a pragmatic approach such a truncation is very appealing,
but the truncation error must be controlled. 
For our simple model, we made a series
of calculations for truncation radii up to $C_p=100$. 
Figure~\ref{fig:rp-dependence} shows the single photo-emission spectrum
for the first excited ionic channel, a diagonal lineout $E_1=E_2$
of  the double-emission spectrum, and two lineouts where one energy energy
is fixed at $E_1=0.1\,au (2.7\,eV)$ and $E_1=0.5\,au (13.6\,eV)$. 
The single electron spectra are hardly affected by the truncation.
The strongest effect, as to be expected, is along the line $E_1=E_2$. 
With $C_p=20$ qualitative differences in the two-electron plots
can be observed, while good qualitative agreement can be
found for $C_p=50$. The values $C_p$ where results are
acceptable must be chosen depending on the system and on 
accuracy requirements. In our case, the on-diagonal spectra are still incorrect 
up to $C_p=50$. In contrast, convergence for larger 
momentum differences appears quite acceptable at $C_p\approx 50$.

\begin{figure}
\begin{center}
\setlength{\unitlength}{1.2mm}
\begin{picture}(35,35)
  \put(-6,33){$|\vr_2|$}
  \put(-6,14){$R_c$}
  \put(0,0){\line(0,1){35}}
  \put(4,26){$\overline{S}$}
  \put( 0,15){\line(1,0){10}}
  \put(10,15){\line(0,1){20}}
  \put(10,15){\line(1,-1){5}}
  \put(4,4){$B$}
  \put(-3,-3){0}
  \put(20,20){$D$}
  \put(15,10){\line(1,0){20}}
  \put(15, 0){\line(0,1){10}}
  \put(26,4){$S$}
  \put(0,0){\line(1,0){35}}
  \put(14,-3){$R_c$}
  \put(35,-3){$|\vr_1|$}

  \put( 13,4){\vector(1,0){5}}
  \put( 13,6){\vector(1,0){5}}
  \put(24,8){\vector(0,1){5}}
  \put(28,8){\vector(0,1){5}}
  \put(32,8){\vector(0,1){5}}

  \put(4,13){\vector(0,1){5}}
  \put(6,13){\vector(0,1){5}}
  \put(8,24){\vector(1,0){5}}
  \put(8,28){\vector(1,0){5}}
  \put(8,32){\vector(1,0){5}}

  \put(10,12){\vector(1,1){4}}
  \put(12,10){\vector(1,1){4}}

\end{picture}\vspace*{2ex}
\caption{\label{fig:partition2}
Integrating the flux for non-vanishing e-e interactions. 
}
\end{center}
\end{figure}
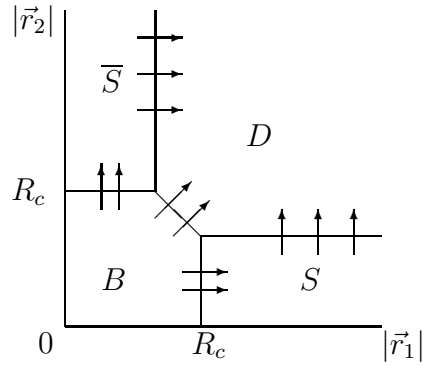

Alternatively to increasing $C_p$, one may 
use a different arrangement of surfaces, where an additional 
surface is placed at $|\vr_1+\vr_2|=R_c$
(see Figure \ref{fig:partition2}).
In scaled center-of-mass and interparticle coordinates $\vr_\pm:=(\vr_1+\vr_2)/\sqrt{2}$
the asymptotic Hamiltonian $H_D(t)$
in region $D$ separates into the center-of-mass motion with a 
Volkov solution and field-free relative motion in the repulsive potential of the electrons
\beq
H_D(t)=\frac12 \left[-i\nabla_+ - \sqrt{2}\vec{A}(t)\right]^2 - \frac12\Delta_- +V_{ee}(\sqrt{2}|\vr_-|), 
\eeq
where $\nabla_+$ and $\Delta_-$ denote the Nabla and Laplace operators in coordinates $\vr_+$ and $\vr_-$,
respectively.
Thus the asymptotic problem reduces to field-free potential scattering and free motion in the laser field.
The contribution going from region $B$ directly into $D$ can be conveniently expanded
in this form. We leave the technical discussion and numerical demonstration of this procedure 
for future work.

\section{Discussion and conclusions}

With the reduction of the simulation volume afforded by t-SURFF, fully dimensional 
calculations of double photo-ionization for a broad range of wave-length and intensities
has come into reach. Scaling to three dimensions depends on the laser frequency:
in the perturbative regime, the number of angular momenta required may remain
as low as 3 \cite{feist08:helium}, which, in linear polarization, 
scales the problem size by $\sim3^3=27$. In that case the 
computational effort is less than one order of magnitude larger than
the calculations presented in section~\ref{sec:xuv}, if one considers that 
the positive and negative half-axes of the one-dimensional model lead to 
a 4-fold scaling of the problem compared to a radial problem. At infrared wave length,
the complete angular phase-space becomes activated. At 800 nm wave length and  
intensities of $\sim 2\times10^{14}W/cm^2$ $\approx30$ angular momenta
are needed for full convergence of electron spectra. For two-electron spectra,
this entails an increase in problem size by  $\sim 30^3/4\approx 7000$ compared to 
the present calculation, if we keep the same standards of accuracy. 
Considering that a typical solution for our problems takes 2 hours on a single CPU, 
one sees that 3D calculation are quiet feasible on a moderate size parallel computer. 
The fact that in the 3D
case the fraction of the phase space, where electron repulsion is important, is much 
smaller than in 1-d, may possibly be exploited for further reduction of the problem size.

The development of t-SURFF is motivated by the photo-ionization problem. What is special
about that problem is that in dipole approximation the time-dependent field affects the 
whole wave function, including the asymptotic region and therefore final momenta cannot
be determined before the end of the pulse, unless one uses knowledge about the time-dependence
of the asymptotic solution. However, also in situations without 
time-dependence of the asymptotic Hamiltonian, the method may turn out to be useful for obtaining
fully differential momentum spectra. All it requires is a reliable absorption method and
knowledge of the solution in the asymptotic region. Among the candidates for further application 
are reactive scattering and chemical break-up processes.

\ack
This work was supported by the ``Munich Advanced Photonics --- MAP'' excellence cluster.

\section*{References}
\bibliographystyle{unsrt}
\bibliography{/home/scrinzi/bibliography/photonics_theory}

\end{document}